\documentclass[twocolumn,superscriptaddress,showpacs,pra]{revtex4}

\usepackage[dvipdfmx]{graphicx}
\usepackage{delarray}
\usepackage{amsmath}
\usepackage{bm}
\usepackage[dvipsnames]{color}
\usepackage{amsfonts}

\newcommand{\ket}[1]{| { #1} \rangle}
\newcommand{\bra}[1]{ \langle {#1} |}

\begin{document}
	\title{Simple security proof of twin-field type quantum key distribution protocol}
	
\author{Marcos Curty}
\affiliation{Escuela de Ingenier\'ia de Telecomunicaci\'on, Dept. of Signal Theory and Communications, University of Vigo, E-36310 Vigo, Spain}
\author{Koji Azuma}
\email{azuma.koji@lab.ntt.co.jp}
\affiliation{NTT Basic Research Laboratories, NTT Corporation, 3-1 Morinosato Wakamiya, Atsugi, Kanagawa 243-0198, Japan}
\affiliation{NTT Research Center for Theoretical Quantum Physics, NTT Corporation, 3-1 Morinosato Wakamiya, Atsugi, Kanagawa 243-0198, Japan}
\author{Hoi-Kwong Lo}
\affiliation{Center for Quantum Information and Quantum Control, Department of Electrical \& Computer Engineering and Department of Physics, University of Toronto, Toronto, Ontario, M5S 3G4, Canada}
\
\date{\today}

\begin{abstract}
Twin-field (TF) quantum key distribution (QKD) was conjectured to beat the private capacity of a point-to-point QKD link by using single-photon interference in a central measuring station. This remarkable conjecture has recently triggered an intense research activity to prove its security. Here, we introduce a TF-type QKD protocol which is conceptually simpler than the original proposal. It relies on local phase randomization, instead of global phase randomization, which significantly simplifies its security analysis and is arguably less demanding experimentally. We demonstrate that the secure key rate of our protocol has a square-root improvement over the point-to-point private capacity, as conjectured by the original TF-QKD scheme. 
\end{abstract}
\maketitle


There is a tremendous research interest towards developing a global quantum internet \cite{K08,AML16,AK17,P16,R18,BA17}, as this could
enable many useful applications of quantum technologies, including, for example,
quantum key distribution (QKD)~\cite{qkd1,LCT14}, blind quantum computing~\cite{blindQ1,blindQ2}, distributed quantum metrology \cite{GJC12,K14} and distributed quantum
computing~\cite{distributedQ}. Among these applications, QKD is certainty the most mature technology today.
Experimentally, long-distance QKD has already been performed over 404~km~\cite{Y16} of telecom fibers
with a standard measurement-device-independent QKD (MDI-QKD) protocol~\cite{LCQ12}, as well as over 1000~km
of free space through satellite to ground links~\cite{L17,Take17}. 
Nonetheless, optical loss in telecom fibers (typically about 0.2~dB/km) poses an important limit to the distance of secure 
QKD without trusted or quantum repeater nodes \cite{BDCZ98,DLCZ01,SSRG09,J09,MSDHN12,ATKI12,ATL15}. Indeed, even with a GHz repetition rate, it would take about 100~years to send a single photon successfully
over 1000~km of a telecom fiber \cite{SSRG09}.
Besides, fundamental limits for the key rate vs distance for secure point-to-point QKD have been obtained recently~~\cite{TGW14,PLOB17}. They essentially
state that, 
in the absence of the repeater nodes, the key rate scales as $\eta$, where $\eta$ is the transmittance of the
channel between Alice and Bob.

Remarkably, Lucamarini {\it et al.}~\cite{LYDS18} have recently proposed a novel MDI-QKD type protocol, called twin-field (TF) QKD, that uses a simple measurement setup for single-photon interference at a central station and is conjectured to beat the fundamental bounds in~\cite{TGW14,PLOB17}, similarly to the provably secure MDI-QKD schemes introduced in~\cite{ATM15,AKB14,PRML14} based on two-photon interference in an adaptive manner. One experimental drawback of TF-QKD is however that single-photon interference requires subwavelength-order phase stability for optical channels, which is more demanding than achieving two-photon interference~\cite{Z07}. Nonetheless, if the conjecture on the security of TF-QKD is proven to be correct, the simplicity and conceptual importance of this protocol will definitively stimulate further investigations. Indeed, very recently, two proofs of security of TF-QKD have been proposed~\cite{TLWL18,MZZ18}; however, none of them is entirely satisfactory. They are rather complicated and require a post-selection on the matching of the global phase of Alice and Bob. This leads to nearly an order of magnitude of drop in the secret key rate.

In this paper, we devise a modified TF-QKD protocol and provide a simple proof of its information-theoretic security. Our proof removes the requirement of post-selection on the matching of the global phase, thus simplifying the proof and elucidating the concepts behind its security. We draw inspiration from quantum repeaters and connect the security of TF-QKD to the study of quantum repeaters. Our proof has also practical impact as it can deliver nearly an {\it order of magnitude} higher secret key rate, compared to the two previous proofs~\cite{ML_QCRYPT18}. For this, we invoke a ``complementarity'' \cite{K09} between the ``phase'' and the ``number'' of a bosonic mode. In particular, to prove the security of a bit encoded in the phase value, we consider what happens if Alice and Bob send optical pulses in number states to the central station. Importantly, the statistics related to this scenario can be estimated by using the decoy-state method \cite{H03,LMC05,W05}. As a result, our protocol can use only {\it local} phase randomization, without the necessity of the global-phase matching condition.

The key idea originates from entanglement generation protocols~\cite{DLCZ01,ATKI12} based on single-photon interference in quantum repeaters. In particular, suppose that Alice and Bob are separated over a distance $L$ and there is a station $C$ right in the middle between them. This central station is connected to Alice (Bob) through an optical fiber with transmittance $\sqrt{\eta}$. If Alice and Bob implement the original MDI-QKD scheme in this scenario, it is clear that the key rate cannot scale better than $\eta$, as this protocol requires that
two photon coincidence events with one photon from Alice and one from Bob interfere in the node $C$. In comparison, TF-QKD can provide a key rate scaling with $\sqrt{\eta}$ because it only requires
singles, {\it i.e.}, one photon (either from Alice or from Bob) reaches the node $C$. Indeed, this scaling improvement is well-known in the field of quantum repeaters. For instance, the performance of the repeater schemes introduced in~\cite{DLCZ01,ATKI12}  scales as $\sqrt{\eta}$ essentially because they use entanglement generation protocols based on single-photon interference in node $C$. Our starting point is then an ideal version of these entanglement generation protocols with an idealized photon source.

{\it Protocol 1:} It consists of the following six steps. (i)~Alice (Bob) first prepares an optical pulse $a$ ($b$) in an entangled state $\ket{\phi_q}_{Aa}=\sqrt{q} \ket{0}_A \ket{0}_a +\sqrt{1-q} \ket{1}_A \ket{1}_a$ ($\ket{\phi_q}_{Bb}$) with $0 \le q \le 1$, where $\ket{0}_{a(b)}$ is the vacuum state and $\ket{1}_{a(b)}$ is the single-photon state for optical pulse $a$ ($b$), and system $A$ ($B$) denotes a qubit in Alice's (Bob's) hands with computational basis $\{\ket{0}_{A(B)},\ket{1}_{A(B)}\}$. (ii) Next, Alice and Bob send the optical pulses $a$ and $b$ through optical channels with transmittance $\sqrt{\eta}$, respectively, to the middle node $C$ in a synchronized manner. (iii) The node $C$ applies to the incoming pulses a 50:50 beamsplitter, followed by two threshold detectors. Let  $D_c$ ($D_d$) denote the detector located at the output port $c$ ($d$) of the beamsplitter associated to constructive (destructive) interference. (iv) The node $C$ announces the measurement outcome $k_c$ ($k_d$) corresponding to detector $D_c$ ($D_d$), where $k_c=0$ and $k_c=1$ ($k_d=0$ and $k_d=1$) indicates a no-click event and a click event, respectively. (v) With probability $p_X$ Alice (Bob) performs the $X$-basis measurement on the qubit $A$ ($B$), and with probability $p_Z$ she (he) performs the $Z$-basis measurement. As a result, Alice (Bob) obtains the bit value $b_A$ ($b_B$), where $(-1)^{b_A} = x$ ($(-1)^{b_B} = x$) for the eigenvalues $x=\pm 1$ of the Pauli operators $\hat{X}$ and $\hat{Z}$. (vi) When node $C$ reports $k_c=1$ and $k_d=0$ ($k_c=0$ and $k_d=1$) and Alice and Bob measure their qubits in the $X$ basis,
$b_A$ and $b_B$ ($b_A$ and $b_B \oplus 1$) are regarded as their raw key.
Note that in this protocol no phase randomization is applied 

We remark that step (iii) above actually corresponds to performing a ``swap test'' on the incoming signals. 
Such a swap test is commonly used in, for example, quantum digital signature schemes~\cite{GC01}
and quantum fingerprinting protocols~\cite{AL14,F15,G16}.

For simplicity and for the moment, let us neglect the effect of the dark counts in the detectors $D_c$ and $D_d$ and assume that  their detection efficiency is perfect. Then, it is straightforward to show that 
the probability $r$ with which node $C$ observes only one click in say detector $D_c$ ($D_d$) in step (iv) above is $r=r_1+r_2$, where
\begin{align}
r_1=& \sqrt{\eta} (1-q) q+(1-q)^2\sqrt{\eta} (1-\sqrt{\eta}),\\
r_2=& \frac{1}{2}(1-q)^2\eta. \label{eq:r}
\end{align}
That is, $r_1$ ($r_2$) corresponds to a detection event  produced by a single-photon (two-photon) pulse. 

Given only one detection click in say 
detector $D_c$ ($D_d$), the joint state of Alice and Bob's qubit systems $A$ and $B$ is denoted by $\hat{\rho}^+_{AB}$ ($\hat{\rho}^-_{AB}$), where
\begin{eqnarray}
&&\hat{\rho}^{\pm}_{AB}=\frac{r_1}{r} \bigg[ \frac{q}{q+(1-q)(1-\sqrt{\eta})} \ket{\Psi^\pm}\bra{\Psi^\pm}_{AB}\nonumber \\
&&+ \frac{(1-q)(1-\sqrt{\eta})}{q+(1-q)(1-\sqrt{\eta})} \ket{11}\bra{11}_{AB} \bigg]
+ \frac{r_2}{r} \ket{11}\bra{11}_{AB},\quad\ \label{eq:state}
\end{eqnarray}
with $\ket{\Psi^\pm}_{AB}:= (\ket{01}_{AB} \pm \ket{10}_{AB})/\sqrt{2}$.

According to Protocol 1, 
the bit-error rate, $e_X$, is defined by the probability with which Alice's and Bob's $X$-basis measurement outcomes are different ({\it i.e.}, $b_A \neq b_B$) when $k_c=1$ and $k_d=0$, or they are equal ($b_A=b_B$) when $k_c=0$ and $k_d=1$. On the other hand, the phase-error rate, $e_Z$, is defined by the probability with which Alice's and Bob's measurement outcomes in the $Z$ basis coincide ($b_A=b_B$) when $k_c+k_d=1$. From Eq.~(\ref{eq:state}) we obtain that $e_X$ and $e_Z$ satisfy
\begin{align}
2 e_X= e_Z= \frac{r_1}{r} \frac{(1-q)(1-\sqrt{\eta})}{q+(1-q)(1-\sqrt{\eta})} + \frac{r_2}{r}. \label{eq:error}
\end{align}
The asymptotic key rate formula $R_X$ is then given by 
\begin{equation}
R_X= 2  r  [1-h(e_X)-h(e_Z)],
\end{equation}
where $2r$ represents the total success probability and $h(x)$ is the binary entropy function, {\it i.e.}, $h(x):=-x \log_2 x -(1-x) \log_2 (1-x)$. The parameter $q$ is chosen such that $R_X$ is maximized for each given distance. 

{\it Protocol 2:} We can also consider a prepare-and-measure version of Protocol 1. For this, we note that, without loss of generality, the measurement in step (v) of Protocol 1 can be done soon after its step (i).
This is because this measurement operation {\it commutes} with all the operations performed in the other steps.
So, the ordering of the steps is not relevant to the physics.
Hence, Protocol 1 is mathematically equivalent to a prepare-and-measure protocol where one omits step (v) and replaces step (i) with the following step: (i') Alice (Bob) prepares an optical pulse $a$ ($b$) in the state $\ket{X_0}_{a(b)}:=\sqrt{q} \ket{0}_{a(b)} + \sqrt{1-q} \ket{1}_{a(b)}$ for $b_A=0$ ($b_B=0$) or in the state $\ket{X_1}_{a(b)}:=\sqrt{q} \ket{0}_{a(b)} - \sqrt{1-q} \ket{1}_{a(b)}$ for $b_A=1$ ($b_B=1$) at random when she (he) chooses the $X$ basis with probability $p_X$, while Alice (Bob) prepares the optical pulse $a$ ($b$) in the state $\ket{Z_0}_{a(b)}:=\ket{0}_{a(b)}$ for $b_A=0$ ($b_B=0$) with probability $q$ or in the state $\ket{Z_1}_{a(b)}:=\ket{1}_{a(b)}$ for $b_A=1$ ($b_B=1$) with probability $1-q$ when she (he) chooses the $Z$ basis with probability $p_Z$. That is, Protocol 2 is composed of step (i'), as well as steps (ii)-(iv) and (vi) from Protocol 1.

\begin{figure}[tb]
  \scalebox{0.25}{\includegraphics{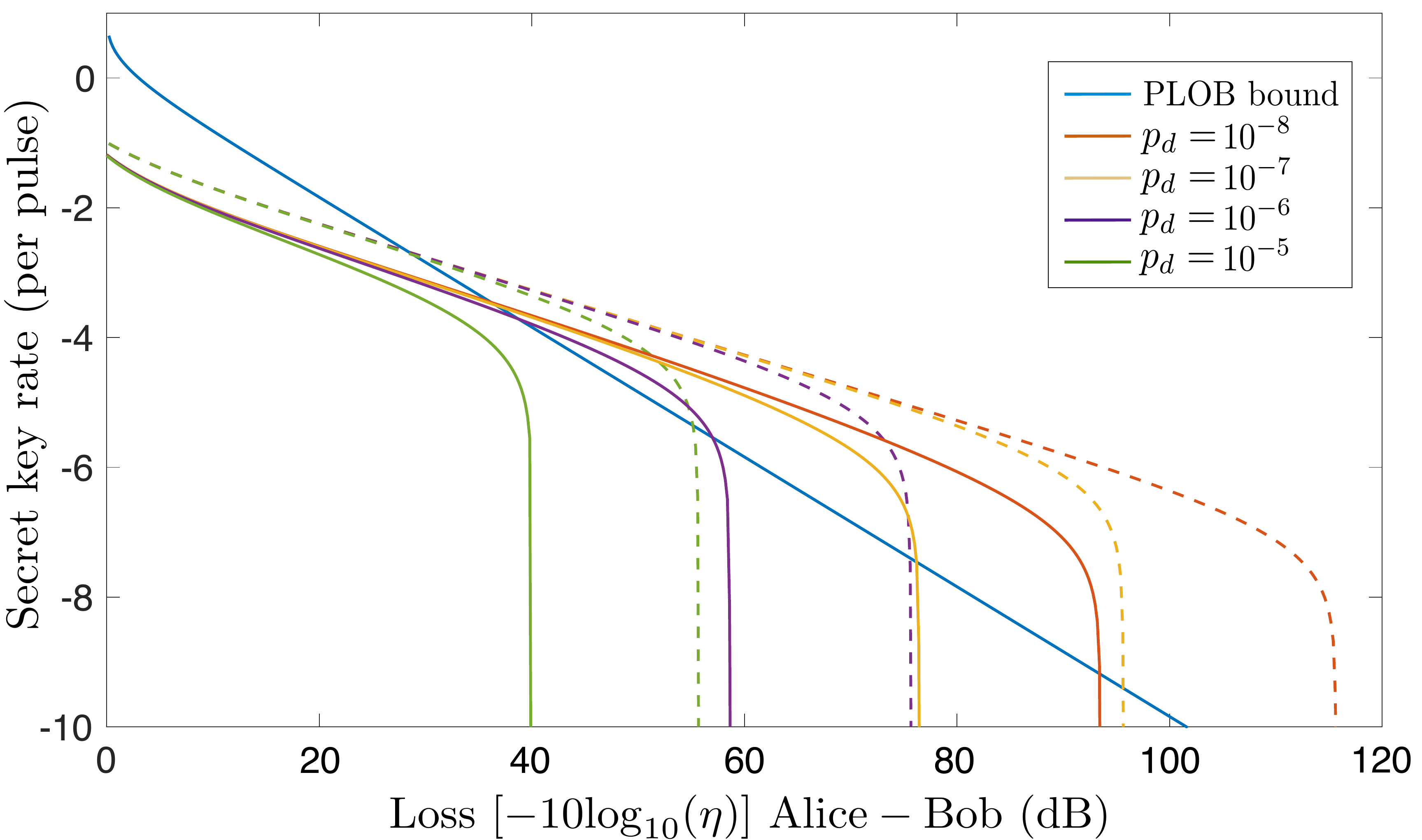}}
  \caption{Secret key rate (per pulse) in logarithmic scale as a function of the {\it overall} loss between Alice and Bob, which includes the finite detection efficiency of the threshold detectors in node $C$. For simulation purposes, we set a misalignment of $2\%$ in each channel Alice-$C$ and Bob-$C$. The dashed (solid) lines correspond to Protocol 1/Protocol 2 (Protocol 3) for different dark count rates, $p_d$, of the detectors in node $C$. The solid blue line illustrates the PLOB bound introduced in~\cite{PLOB17}.
  Our simulation results show clearly that, even
  in the presence of reasonably low values of dark counts
  of about $10^{-6}$ per pulse and misalignment, the Protocols could beat the PLOB bound.	
    \label{fig:graph-pra}}
\end{figure}

In Fig.~\ref{fig:graph-pra}, we show the performance of these two protocols by maximizing $R_X$ over $q$ as a function of the overall loss between Alice and Bob. According to our computation calculation, the optimal value of $q=||{}_a\bra{0}\ket{\phi_q}_{Aa}||^2$ starts from about 0.88 at $0$~dB, and then monotonically increases with the loss up to a value of about 0.94 at $20$~dB, and afterward remains basically constant. The high value of $q$ suggests that the states $\ket{X_k}$ ($k=0,1$) could be replaced by coherent states $\ket{(-1)^k \alpha} $ by choosing an appropriate amplitude $\alpha(> 0)$, as their good approximation. Also, since the states $\ket{Z_k}$ ($k=0,1$) are number states, Alice and Bob could estimate the phase-error rate $e_Z$ by using phase-randomized coherent states in combination with the decoy-state method. These two observations lead to the following practical protocol.

{\it Protocol~3:} It is composed of the following modified first step (i'') together with steps (ii)-(iv) and (vi) from Protocol~1:
(i'') Alice (Bob) first chooses the $X$ basis with probability $p_X$ and the $Z$ basis with probability $p_Z$. If her (his) choice is the $X$ basis, she (he) prepares an optical pulse $a$ ($b$) in a coherent state $\ket{\alpha}_{a(b)}$ for $b_A=0$ ($b_B=0$) or $\ket{-\alpha}_{a(b)}$ for $b_A=1$ ($b_B=1$) at random. 
If her (his) choice is the $Z$ basis, she (he) prepares an optical pulse $a$ ($b$) in a phase-randomized coherent state $\hat{\rho}_{a, \beta_A}$ ($\hat{\rho}_{b, \beta_B}$) whose amplitude $\beta_A$ ($\beta_B$) is chosen from a set $S=\{\beta_i \}_i$ of real nonnegative numbers $\beta_i \ge 0$ at random. 

It is important to note that Protocol~3 requires synchronization of phase references for Alice and Bob. However, since in QKD Alice and Bob may use ancillary strong pulses generated by lasers to
establish such a pulse reference, we believe that establishing the phase references is practical. In this scenario, we assume that all the $X$-basis (key generation) states of Alice and Bob are either of the same or opposite phase. That is, {\it no} phase randomization is needed for the key generation states. 
In contrast, all the $Z$-basis states (used for test for tampering) of Alice and Bob have
random phases, which allows us to apply the decoy state technique
to these states to infer the contributions from the vacuum, single-photon, and multi-photon components. 
Also, note that $p_X$ can be chosen much higher than $p_Z$ to have a high key generation rate.

{\it Security proof of Protocol~3:} For simplicity we shall consider the asymptotic scenario where Alice and Bob emit an infinite number of signals. Also, without loss of generality, we shall assume that the node $C$ is under the full control of an eavesdropper, Eve. After a QKD run, Alice and Bob can estimate the probability distribution $p_{ZZ}(k_c,k_d|\beta_A,\beta_B)$ ($p_{XX}(k_c,k_d|b_A,b_B)$) over $k_c$ and $k_d$ given the choice of $\beta_A$ and $\beta_B$ ($b_A$ and $b_B$) and the selection of the $Z$ ($X$) basis. By noting that
\begin{equation}
p_{XX} (b_A,b_B|k_c,k_d) = \frac{1}{4}  \frac{p_{XX}(k_c,k_d|b_A,b_B)}{p_{XX}(k_c,k_d)},
\end{equation}
where
\begin{equation}
p_{XX}(k_c,k_d)=\frac{1}{4}\sum_{b_A,b_B=0,1} p_{XX}(k_c,k_d|b_A,b_B),
\end{equation}
we have that the bit-error rate, $e_{X,k_ck_d}$, for Eve's announcement of $k_c$ and $k_d$ is defined by 
\begin{eqnarray}
e_{X,10} &=& \sum_{i,j|i\oplus j =1}  p_{XX} (b_A=i,b_B=j|k_c=1,k_d=0), \nonumber \\
e_{X,01} &=& \sum_{j=0,1}  p_{XX} (b_A=j,b_B=j|k_c=0,k_d=1).
\end{eqnarray}

Next we consider the decoy-state method. In particular, since when Alice and Bob choose the $Z$ basis in step (i'') of Protocol 3 they prepare phase-randomized coherent states, Eve cannot distinguish this step from the following fictitious scenario: Alice (Bob) prepares an optical pulse $a$ ($b$) in a number state $\ket{n_A}_a$ ($\ket{n_B}_b$) according to a Poissonian distribution $P_{\beta_A^2}(n_A)$ ($P_{\beta_B^2}(n_B)$), where $P_\lambda(n)= (e^{-\lambda}\lambda^n)/n!$. In this fictitious scenario, Eve's attack can only depend on the number states $\ket{n_A}$ and $\ket{n_B}$. This implies that Eve's announcement of $k_c$ and $k_d$ follows a probability distribution $p_{ZZ}(k_c,k_d|n_A,n_B)$.
Then, we have
\begin{multline}
p_{ZZ}(k_c,k_d|\beta_A,\beta_B)  \\
= \sum_{n_A,n_B=0}^\infty p_{ZZ}(k_c,k_d|n_A,n_B) P_{\beta_A^2}(n_A) P_{\beta_B^2}(n_B)
\end{multline}
for any $\beta_A$ and $\beta_B$. That is, once Alice and Bob know $p_{ZZ}(k_c,k_d|\beta_A,\beta_B)$ for any $\beta_A$ and $\beta_B$, they can use the decoy-state method to estimate $p_{ZZ}(k_c,k_d|n_A,n_B)$ based on their knowledge of $P_{\beta_A^2}(n_A)$ and $P_{\beta_B^2}(n_B)$.

The next step is to relate the conditional probabilities $p_{ZZ}(k_c,k_d|n_A,n_B)$ with the phase-error rate to prove security~\cite{K09}. For this, note that if Alice and Bob choose the $X$ basis in step (i'') of Protocol~3, Eve cannot distinguish this step from the following fictitious step: Alice (Bob) prepares an optical pulse $a$ ($b$) and a qubit $A$ ($B$) in
an entangled state $\ket{\psi_X}_{Aa}=(\ket{+}_A \ket{\alpha}_a +\ket{-}_A\ket{- \alpha}_a)/\sqrt{2}$
($\ket{\psi_X}_{Bb}$) with $\ket{\pm}_{A(B)}:=(\ket{0}_{A(B)}\pm \ket{1}_{A(B)})/\sqrt{2}$. By running this fictitious step together with steps (ii)-(iv) in order, Alice and Bob obtain a state 
\begin{equation}
\ket{\chi_{k_c k_d}}_{Aa'Bb'}:=\frac{\hat{M}^{ab}_{k_c k_d} \ket{\psi_X}_{Aa} \ket{\psi_X}_{Bb}}{\sqrt{p_{XX}(k_c,k_d)} }, \label{eq:chi}
\end{equation}
with probability $p_{XX}(k_c,k_d)$, where $\hat{M}^{ab}_{k_c k_d}$ is the Kraus operator corresponding to the announcement of $k_c$ and $k_d$.
The phase-error rate, $e_{Z,k_c k_d}$, is then defined by
\begin{equation}
e_{Z,k_c k_d} = \sum_{j =0,1}\| {}_{AB}\bra{jj}\ket{\chi_{k_c k_d}}_{Aa'Bb'} \|^2 . \label{eq:p-error}
\end{equation}
Since ${}_A \bra{i}\ket{\psi_X}_{Aa}= \ket{C_i}_a$ with unnormalized cat states
\begin{align}
 \ket{C_0}_a = & e^{-\frac{\alpha^2}{2}} \sum_{n=0}^\infty \frac{\alpha^{2n}}{\sqrt{(2n)!}} \ket{2n}_a=: \sum_{n=0}^\infty c_n^{(0)}  \ket{n}_a,\\
 \ket{C_1}_a = & e^{-\frac{\alpha^2}{2}}  \sum_{n=0}^\infty \frac{\alpha^{2n+1}}{\sqrt{(2n+1)!}} \ket{2n+1}_a =: \sum_{n=0}^\infty c_n^{(1)}  \ket{n}_a,
\end{align}
for nonnegative coefficients $c_n^{(i)}\ge 0$,
from Eq.~(\ref{eq:chi}) and for any $i,j=0,1$, we have
\begin{multline}
p_{XX}(k_c,k_d)  \| {}_{AB}\bra{ij}\ket{\chi_{k_c k_d}}_{Aa'Bb'} \|^2 \\
={}_a \bra{C_i} {}_b \bra{C_j} (\hat{M}^{ab}_{k_c k_d})^\dag  \hat{M}^{ab}_{k_c k_d} \ket{C_i}_a\ket{C_j}_b \\
= \sum_{m_A,m_B,n_A,n_B=0}^\infty c_{m_A}^{(i)}  c_{m_B}^{(j)}   c_{n_A}^{(i)}  c_{n_B}^{(j)} \\
\times {}_a\bra{m_A} {}_b\bra{m_B} (\hat{M}^{ab}_{k_c k_d})^\dag  \hat{M}^{ab}_{k_c k_d} \ket{n_A}_a\ket{n_B}_b  \\
\le \sum_{m_A,m_B,n_A,n_B=0}^\infty c_{m_A}^{(i)}  c_{m_B}^{(j)}   c_{n_A}^{(i)}  c_{n_B}^{(j)} \\
\times \| \hat{M}^{ab}_{k_c k_d} \ket{m_A}_a\ket{m_B}_b \|   \| \hat{M}^{ab}_{k_c k_d} \ket{n_A}_a\ket{n_B}_b \| \\
= \left[\sum_{n_A,n_B=0}^\infty    c_{n_A}^{(i)}  c_{n_B}^{(j)}  \sqrt{p_{ZZ}(k_c,k_d|n_A,n_B)} \right]^2,
\end{multline}
where we have used the Cauchy-Schwarz inequality and $\| \hat{M}^{ab}_{k_c k_d} \ket{m_A}_a\ket{m_B}_b \|^2 =p_{ZZ}(k_c,k_d|m_A,m_B)$. 
By combining these results with Eq.~(\ref{eq:p-error}),
we conclude 
\begin{multline}
p_{XX}(k_c,k_d)  e_{Z,k_c k_d} \le p_{XX}(k_c,k_d)  e^{\rm upp}_{Z,k_c k_d} \\ 
:=  \sum_{j=0,1} \left[\sum_{n_A,n_B=0}^\infty    c_{n_A}^{(j)}  c_{n_B}^{(j)}  \sqrt{p_{ZZ}(k_c,k_d|n_A,n_B)} \right]^2. \label{eq:key}
\end{multline}
That is, we can estimate an upper bound, $e^{\rm upp}_{Z,k_c k_d}$, on the phase-error rate $e_{Z,k_c k_d} $ from the observed data.
This means that the asymptotic key rate formula, $R_{X,k_ck_d}$, can be lower bounded as
\begin{align}
&R_{X,k_ck_d} = p_{XX}(k_c,k_d) \left[1- h(e_{X,k_ck_d}) - h(e_{Z,k_ck_d})\right]  \nonumber \\
&\ge   p_{XX}(k_c,k_d) [1- h(e_{X,k_ck_d}) - h(\min\{1/2,e^{\rm upp}_{Z,k_ck_d}\})] \nonumber  \\
&=:R^{\rm low}_{X,k_ck_d}, 
\end{align}
which leads to the final key rate formula:
\begin{equation}
R_X = R_{X,10}+R_{X,01} \ge R^{\rm low}_{X,10} +R^{\rm low}_{X,01}=:R^{\rm low}_{X}.
\end{equation}

The performance of Protocol~3 is illustrated in Fig.~\ref{fig:graph-pra}, where we maximize a further lower bound on $R^{\rm low}_{X}$ over $\alpha$ as a function of the overall loss between Alice and Bob. In particular, here we assume the asymptotic scenario where Alice and Bob use an infinite number of decoy settings and they can estimate the probabilities $p_{ZZ}(k_c,k_d|n_A,n_B)$, with $(n_A, n_B)= (0,0), (0,2), (2,0), (2,2),(1,1),(1,3),(3,1)$, precisely, while the remaining probabilities are simply upper bounded as $p_{ZZ}(k_c,k_d|n_A,n_B)\le 1$ (although, clearly, the more probabilities $\{p_{ZZ}(k_c,k_d|n_A,n_B)\}_{n_A,n_B}$ Alice and Bob tightly estimate, the higher the resulting key rate is).  Importantly, Fig.~\ref{fig:graph-pra} demonstrates that $R^{\rm low}_{X}$ has $\sqrt{\eta}$ scaling. In the Appendix, it is also confirmed
 that the use of three decoy states (that is, setting $S=\{\beta_i \}_{i=1,2,3}$ in Protocol~3), rather than infinite decoy states, is enough for Protocol~3 to achieve a similar performance to Fig.~1. Besides, remarkably, Protocol~3 is quite robust against phase mismatch between Alice-$C$ and Bob-$C$ channels. See Appendix for the details.

The fact that the cases $(n_A, n_B)= (0,1)$ or $(1,0)$
	do not contribute at all to the phase-error rate is remarkable. The reason for this behaviour is the following.
	The even (odd) cat state corresponding to $j=0$ ($j=1$) in Eq. (12) (Eq.~(13)) includes only even (odd) photons. And Eq. (14) considers what happens when Alice's input and Bob's input are both (phase-randomized) even cat states or both (phase-randomized) odd cat states. Thus, the terms $(0,1)$ and $(1,0)$ never contribute. This means that by lowering bounding other contributions (such as $(n_A, n_B)= (0,0), (0,2), (2,0),\ldots$) with decoy states,
	one can severely limit the amount of information Eve has on the
	sifted key. Moreover, note that the signals contain mainly only one photon or less 
	originating from {\it either} Alice or Bob.
	The net transmittance of the signal is thus of order $\sqrt{\eta}$, which leads to a very high key rate for TF-type QKD at long distances.
	That is, it is mainly the interference between the single photon component
	generated by either Alice or Bob that leads to security.

We also remark that by regarding $c^{(j)}_n=\sqrt{q_{n|j}q_j}$ with probability distributions $\{q_j\}_j$ and $\{q_{n|j}\}_n$, one can consider the terms $ c_{n_A}^{(j)}  c_{n_B}^{(j)}  \sqrt{p_{ZZ}(k_c,k_d|n_A,n_B)} $ to be the square root of a joint probability. This implies that $p_{XX}(k_c,k_d)  (e_{Z,k_c k_d} -  e^{\rm upp}_{Z,k_c k_d} ) (\le0)$ is a convex function over probabilities that can be obtained by performing positive operator-valued measure (POVM) measurements on a quantum state for a round in a virtual
scenario. This is enough \cite{TLKB09} to prove the security of Protocol~3 against coherent attacks, thanks to Azuma's inequality~\cite{A67}.

Finally, we note that the structure of the security proof of Protocol~3 resembles that for the loss-tolerant QKD protocol~\cite{TCKLA14}. Therefore, its extension to the finite-key scenario could be readily done by using similar techniques like those employed in~\cite{MCLIT15,NMIYIT16,M18}, in combination with the decoy-state analysis employed in standard MDI-QKD~\cite{CXCLTL14}.

In summary, we have introduced a novel TF-type QKD protocol, together with a simple proof of its security, which can beat the fundamental 
bounds on the private capacity of point-to-point QKD over a lossy optical channel presented in~\cite{TGW14,PLOB17}. Its secret key rate 
scales as $\sqrt{\eta}$ rather than $\eta$, being $\eta$ the transmittance of the quantum channel. This protocol could also be regarded
as a phase-encoding MDI-QKD scheme with single-photon interference. Indeed, it inherits the major
advantage of standard MDI-QKD, {\it i.e.}, it is robust against any side channel in the measurement unit. 
 
{\it Acknowledgements.}---We thank G.~Kato and Y.~Zhang for helpful discussions,
M.~Lucamarini and K.~Tamaki for discussions related to the papers~\cite{TLWL18,LYDS18},
and X. Ma and P. Zeng for discussions related to the paper~\cite{MZZ18}. K.A. thanks support, in part, from PRESTO, JST JPMJPR1861. 
M.C. acknowledges support from the Spanish Ministry of Economy and Competitiveness (MINECO), the Fondo Europeo de Desarrollo Regional (FEDER) through grants TEC2014-54898-R and TEC2017-88243-R, and the European Union's Horizon 2020 research and innovation programme under the Marie Sklodowska-Curie grant agreement No 675662 (project QCALL). H.-K.L. thanks the US Office of Naval Research, NSERC, CFI, ORF, MITACS, Huawei Technologies Canada Co., Ltd, and the Royal Bank of Canada for financial support.

{\it Author contributions.} M.C. and K.A. contributed equally to this work; M.C. contributed more to the protocol design and K.A. to its security proof. H.-K.L. triggered the consideration of this
research project. All authors contributed to the writing and generalization of the ideas.

{\it Note added.}---During the preparation of this paper, two other works considering similar protocols have been posted on preprint servers~\cite{C18} or presented in a conference~\cite{L18}.
We thank N.~L\"{u}tkenhaus' group for discussions regarding the results in~\cite{L18}. While our formulation and discussion for security have some similarities with these results, there are also differences in the 
methodology and our initial idea was conceived independently of these two works.

\section*{APPENDIX}

\section*{Evaluation of the secret key rate formula}

From the main text, we have that a lower bound on the secret key rate delivered by Protocol~3 can be written as
\begin{equation}\label{key_rate}
R^{\rm low}_{X}= \max{\{R^{\rm low}_{X,10},0\}} +\max{\{R^{\rm low}_{X,01},0\}},
\end{equation}
where the terms $R^{\rm low}_{X,k_ck_d}$ have the form
\begin{eqnarray}\label{qwe5}
R^{\rm low}_{X,k_ck_d}&=&p_{XX}(k_c,k_d) [1- h(e_{X,k_ck_d}) \nonumber \\
&-& h(\min\{1/2,e^{\rm upp}_{Z,k_ck_d}\})].
\end{eqnarray}
In Eq.~(\ref{qwe5}), the quantity $p_{XX}(k_c,k_d)$ corresponds to the conditional probability that node $C$ announces the measurement outcome $(k_c,k_d)$ given that both Alice and Bob emit a signal encoded in the $X$ basis; see Eq.~(7). The term $e_{X,k_ck_d}$ represents the bit-error rate; it is given by Eq.~(8). Finally, $e^{\rm upp}_{Z,k_ck_d}$ refers to an upper bound on the phase-error rate. This last quantity can be written as
\begin{multline}\label{phase}
e_{Z,k_c k_d}^{\rm upp} = \frac{1}{p_{XX}(k_c,k_d)}\\ \times \sum_{j=0,1} \left[ \sum_{n_A,n_B=0}^\infty c_{n_A}^{(j)} c_{n_B}^{(j)}  \sqrt{p_{ZZ} (k_c,k_d|n_A,n_B)} \right]^2 \\
=\frac{1}{p_{XX}(k_c,k_d)} \sum_{j=0,1} \Biggl[ \sum_{m_A,m_B=0}^\infty c_{2m_A +j}^{(j)} c_{2m_B+j}^{(j)} \\
\times  \sqrt{p_{ZZ} (k_c,k_d|2m_A+j, 2m_B+j)} \Biggr]^2,
\end{multline}
where the coefficients $c_m^{(j)}$, with $m \in \{0,1,2,\ldots \} =: \mathbb{N}_0$ and $j\in \{0,1\}$, are given by Eqs.~(12)-(13). The second equality in Eq.~(\ref{phase}) is due to the fact that these coefficients satisfy $c_{2m+1}^{(0)} =c_{2m}^{(1)}=0$ for any $m \in \mathbb{N}_0$. 

To evaluate Eq.~(\ref{phase}), for each $j\in \{0,1\}$ we estimate upper bounds, $p_{ZZ}^{\rm upp}(k_c,k_d|2m_A+j,2m_B+j)$, on the conditional probabilities $p_{ZZ}(k_c,k_d|2m_A+j,2m_B+j)$ for $(m_A,m_B) \in {\cal S}_j$, where ${\cal S}_j$ is a chosen subset of $\{(m_A,m_B)| m_A,m_B \in \mathbb{N}_0\}$. Those probabilities $p_{ZZ}(k_c,k_d|2m_A+j,2m_B+j)$ with $(m_A,m_B)  \not\in {\cal S}_j$ are trivially upper bounded as $p_{ZZ}(k_c,k_d|2m_A+j,2m_B+j) \le 1$. As a result, we have that $e_{Z,k_c k_d}^{\rm upp} $ is upper bounded by
\begin{multline}\label{phase_used}
e_{Z,k_c k_d}^{\rm upp} \le 
\frac{1}{p_{XX}(k_c,k_d)} \sum_{j=0,1} \Biggl[ \sum_{(m_A,m_B) \in {\cal S}_j} c_{2m_A +j}^{(j)} c_{2m_B+j}^{(j)}  \\ \times  \sqrt{p_{ZZ}^{\rm upp} (k_c,k_d|2m_A+j, 2m_B+j)} + \Delta_j \Biggr]^2 ,
\end{multline}
where the parameters $\Delta_j$ simply refer to the residual terms associated with the probabilities $p_{ZZ}(k_c,k_d|2m_A+j,2m_B+j)$ with $(m_A,m_B)  \not\in {\cal S}_j$. These parameters can be written as

\begin{equation}
 \Delta_j = \sum_{(m_A,m_B) \not\in {\cal S}_j} c^{(j)}_{2m_A+j} c^{(j)}_{2m_B+j}.
\end{equation}

In the next section of this Appendix we present a simple numerical method to estimate the terms $p_{ZZ}^{\rm upp}(k_c,k_d|2m_A+j,2m_B+j)$ for $(m_A,m_B)\in\mathcal{S}_j $ ($j\in \{0,1\}$). Afterward, we introduce a channel model to determine the quantities $p_{XX}(k_c,k_d)$ and $e_{X,k_ck_d}$ and thus be able to evaluate Eq.~(\ref{key_rate}). Finally, in the last section of the Appendix, we provide some additional simulation results that complement those presented in the main text. 

\section*{Estimation of $p_{ZZ}^{\rm upp}(k_c,k_d|2m_A+j,2m_B+j)$}

To estimate the parameters $p_{ZZ}^{\rm upp}(k_c,k_d|2m_A+j,2m_B+j)$ with $(m_A,m_B)  \in {\cal S}_j$ one can use analytical or numerical methods. For simplicity, here we consider a numerical procedure that is valid for any number of decoy intensity settings and for any photon-number distribution of the signals emitted by Alice and Bob. More precisely, we show that this estimation problem can be written as a linear program, which can be solved efficiently in polynomial time~\cite{lp_book}. 

Our starting point is Eq.~($9$), which we reproduce here for completeness, 
\begin{multline}\label{9}
p_{ZZ}(k_c,k_d|\beta_A,\beta_B) \\
= \sum_{n_A,n_B=0}^\infty p_{ZZ}(k_c,k_d|n_A,n_B) P_{\beta_A^2}(n_A) P_{\beta_B^2}(n_B).
\end{multline}
This equation relates the experimentally observed gains, $p_{ZZ}(k_c,k_d|\beta_A,\beta_B)$, with the unknown parameters $p_{ZZ}(k_c,k_d|n_A,n_B)$ through the known photon-number distributions, $P_{\beta_A^2}(n_A)$ and $P_{\beta_B^2}(n_B)$, of Alice and Bob's emitted signals. For example, when Alice and Bob use phase-randomised weak coherent pulses, then $P_\lambda(n)= (e^{-\lambda}\lambda^n)/n!$ with $\lambda$ being the intensity of the light source. 

To formulate the estimation of $p_{ZZ}^{\rm upp}(k_c,k_d|2m_A+j,2m_B+j)$ for given $(m_A,m_B)$ and $j$ as a linear program which can be solved numerically, we first reduce the number of unknown parameters, $p_{ZZ}(k_c,k_d|n_A,n_B)$, in Eq.~(\ref{9}) to a finite set. For this, we derive both a lower and an upper bound for each quantity $p_{ZZ}(k_c,k_d|\beta_A,\beta_B)$. In particular, since the terms $p_{ZZ}(k_c,k_d|n_A,n_B) P_{\beta_A^2}(n_A) P_{\beta_B^2}(n_B)\geq{}0$ for all $n_A,n_B$, from Eq.~(\ref{9}) we have that $p_{ZZ}(k_c,k_d|\beta_A,\beta_B)$ satisfies
\begin{multline}\label{lower}
p_{ZZ}(k_c,k_d|\beta_A,\beta_B) \\
\geq \sum_{(n_A,n_B)\in{\mathcal S}_{\rm cut}} p_{ZZ}(k_c,k_d|n_A,n_B) P_{\beta_A^2}(n_A) P_{\beta_B^2}(n_B),
\end{multline}
where ${\mathcal S}_{\rm cut}$ denotes the finite set of indexes $(n_A,n_B)$ identifying those $p_{ZZ}(k_c,k_d|n_A,n_B)$ which are considered as unknown parameters when estimating $p_{ZZ}^{\rm upp}(k_c,k_d|2m_A+j,2m_B+j)$. The set ${\mathcal S}_{\rm cut}$ is typically larger than the sets ${\cal S}_j$, which indicate the probabilities $p_{ZZ}(k_c,k_d|2m_A+j,2m_B+j)$ which are actually nontrivially upper bounded when evaluating Eq.~(\ref{phase_used}). The only requirement here is that for all $(m_A,m_B)\in {\cal S}_j$ ($j\in \{0,1\}$) then $(2m_A+j,2m_B+j)\in{\mathcal S}_{\rm cut}$. Otherwise, the set of linear equations given by Eq.~(\ref{lower}) (and also given by Eq.~(\ref{upper}) below) would not include as unknowns some of the probabilities $p_{ZZ}(k_c,k_d|2m_A+j,2m_B+j)$ that we wish to upper bound. As a result, the missing probabilities would be trivially upper bounded by $1$. 

Similarly to Eq.~(\ref{lower}), we also have that 
\begin{multline}\label{upper}
p_{ZZ}(k_c,k_d|\beta_A,\beta_B) \\
\leq \sum_{(n_A,n_B)\in{\mathcal S}_{\rm cut}} p_{ZZ}(k_c,k_d|n_A,n_B) P_{\beta_A^2}(n_A) P_{\beta_B^2}(n_B) \\
+\sum_{(n_A,n_B)\notin{\mathcal S}_{\rm cut}} P_{\beta_A^2}(n_A) P_{\beta_B^2}(n_B)\\
= \sum_{(n_A,n_B)\in{\mathcal S}_{\rm cut}} p_{ZZ}(k_c,k_d|n_A,n_B) P_{\beta_A^2}(n_A) P_{\beta_B^2}(n_B) \\
+\left(1-\sum_{(n_A,n_B)\in{\mathcal S}_{\rm cut}} P_{\beta_A^2}(n_A) P_{\beta_B^2}(n_B)\right),
\end{multline}
where in the first inequality we use the fact that $p_{ZZ}(k_c,k_d|n_A,n_B)\leq{}1$ for any $n_A$ and $n_B$, and the equality holds because $\sum_{n_A,n_B=0}^\infty P_{\beta_A^2}(n_A) P_{\beta_B^2}(n_B)=1$.

In so doing, we can write the estimation of each $p_{ZZ}^{\rm upp}(k_c,k_d|2m_A+j,2m_B+j)$, with $(m_A,m_B)  \in {\cal S}_j$, by using the following linear program
\begin{widetext}
\begin{eqnarray}\label{lp}
\max && p_{ZZ}(k_c,k_d|2m_A+j,2m_B+j)\nonumber \\
\textrm{subject to}&& p_{ZZ}(k_c,k_d|\beta_A,\beta_B)\geq \sum_{(n_A,n_B)\in{\mathcal S}_{\rm cut}} p_{ZZ}(k_c,k_d|n_A,n_B) P_{\beta_A^2}(n_A) P_{\beta_B^2}(n_B), \quad \forall\beta_A,\beta_B, \nonumber \\
&& 1- p_{ZZ}(k_c,k_d|\beta_A,\beta_B)\geq \sum_{(n_A,n_B)\in{\mathcal S}_{\rm cut}} \left[1-p_{ZZ}(k_c,k_d|n_A,n_B)\right] P_{\beta_A^2}(n_A) P_{\beta_B^2}(n_B), \quad \forall\beta_A,\beta_B, \nonumber \\
&& 0\leq p_{ZZ}(k_c,k_d|n_A,n_B)\leq 1, \quad \forall (n_A,n_B)\in{\mathcal S}_{\rm cut}.
\end{eqnarray}
\end{widetext}

\section*{Channel model}\label{channel_m}

Here, we introduce a simple channel model to simulate the quantities that would be observed in an actual experiment. We use this channel model to estimate the parameters $p_{XX}(k_c,k_d)$, $e_{X,k_ck_d}$ and $p_{ZZ}(k_c,k_d|\beta_A,\beta_B)$. In all the mathematical expressions within this section we consider only the cases where $k_c\oplus{}k_d=1$.

In particular, we model the overall loss between Alice (Bob) and node $C$ with a beamsplitter of transmittance $\sqrt{\eta}$. This overall loss includes as well the non-unity detection efficiency of the detectors $D_c$ and $D_d$ located at node $C$.  Also, we consider that the two quantum channels connecting Alice and Bob with node $C$ introduce both polarization and phase misalignments. The polarization misalignment of the link Alice-$C$ (Bob-$C$) is modelled with a unitary operation that transforms the polarization input mode, $a_{\rm in}^\dagger$ ($b_{\rm in}^\dagger$), into the orthogonal polarization output modes, $a_{\rm out}^\dagger$ and $a_{{\rm out}\perp}^\dagger$ ($b_{\rm out}^\dagger$ and $b_{{\rm out}\perp}^\dagger$) as follows: $a_{\rm in}^\dagger\rightarrow\cos{\theta_{\rm A}}a_{\rm out}^\dagger-\sin{\theta_{\rm A}}a_{{\rm out}\perp}^\dagger$ ($b_{\rm in}^\dagger\rightarrow\cos{\theta_{\rm B}}b_{\rm out}^\dagger-\sin{\theta_{\rm B}}b_{{\rm out}\perp}^\dagger$) for certain angle $\theta_{\rm A}$ ($\theta_{\rm B}$). To model the phase mismatch between Alice and Bob's signals arriving at node $C$, we shift the phase of say Bob's signals by an angle $\phi=\delta\pi$ for a certain parameter $\delta$. Finally, we assume that the functioning of the detectors $D_c$ and $D_d$ is independent of the polarization of the incoming signals, and they suffer from a dark count probability, $p_d$, which is, to a good approximation, also independent of the signals received and has the same value for both detectors.

In this scenario, it can be shown that the gains ({\it i.e.}, the probabilities to observe a certain detection pattern $(k_c,k_d)$) associated to the signal states in the $X$ basis can be expressed as 
\begin{multline}\label{qwe}
p_{XX}(k_c,k_d|b_A,b_B)= \\
(1-p_d)\bigg[p_de^{-2\gamma}+q_{XX}(k_c,k_d|b_A,b_B)\bigg],
\end{multline}
for all $b_A,b_B=0,1$, and where $\gamma=\sqrt{\eta} \alpha^2$, being $\alpha>0$ the amplitude of the signal states. The term $q_{XX}(k_c,k_d|b_A,b_B)$ that appears in Eq.~(\ref{qwe}) denotes the gain corresponding to the case without dark counts in the detectors. Its expression is given by
\begin{equation}
q_{XX}(k_c,k_d|b_A,b_B)= \left\{ \begin{array}{ll}
 f^-_{(\theta,\phi,\gamma)} & \textrm{if $k_c\oplus{}b_A\oplus{}b_B=1$,}\\
 f^+_{(\theta,\phi,\gamma)} & \textrm{if $k_c\oplus{}b_A\oplus{}b_B=0$,}
\end{array} \right.
\end{equation}
with $\theta=\theta_{\rm A}-\theta_{\rm B}$. The functions $f^\pm_{(\theta,\phi,\gamma)}$ have the form
\begin{equation}
f^\pm_{(\theta,\phi,\gamma)}=e^{-\gamma[1\pm\Omega(\phi,\theta)]}-e^{-2\gamma},
\end{equation}
with $\Omega(\phi,\theta)=\cos{\phi}\cos{\theta}$.

Similarly, we find that the probabilities $p_{XX}(k_c,k_d)$ can be written as
\begin{eqnarray}\label{pxx}
p_{XX}(k_c,k_d)&=&\frac{1}{2}(1-p_d)\left(e^{-\gamma\Omega(\phi,\theta)}+e^{\gamma\Omega(\phi,\theta)}\right)e^{-\gamma} \nonumber \\
&-&(1-p_d)^2e^{-2\gamma}.
\end{eqnarray}

Also, from Eq.~(\ref{qwe}), together with Eqs.~(6)-(8), we find that the bit-error rates, $e_{X,k_ck_d}$, satisfy
\begin{equation}\label{berror}
e_{X,k_ck_d}=\frac{e^{-\gamma\Omega(\phi,\theta)}-(1-p_d)e^{-\gamma}}{e^{-\gamma\Omega(\phi,\theta)}+e^{\gamma\Omega(\phi,\theta)}-2(1-p_d)e^{-\gamma}}.
\end{equation}

Finally, it can be shown that the gains associated to the signals in the $Z$ basis are given by
\begin{multline}\label{qwe2}
p_{ZZ}(k_c,k_d|\beta_A,\beta_B)= \\
(1-p_d)\bigg[p_de^{-(\beta_A^2+\beta_B^2)\sqrt{\eta}}+q_{ZZ}(k_c,k_d|\beta_A,\beta_B)\bigg],
\end{multline}
where $q_{ZZ}(k_c,k_d|\beta_A,\beta_B)$ denotes again the corresponding gain assuming no dark counts in the detectors. Its expression is
\begin{eqnarray}
q_{ZZ}(k_c,k_d|\beta_A,\beta_B)&=&e^{-\frac{(\beta_A^2+\beta_B^2)\sqrt{\eta}}{2}}I_{0}\left(\beta_A\beta_B\sqrt{\eta}\cos{\theta}\right)\nonumber \\
&-&e^{-(\beta_A^2+\beta_B^2)\sqrt{\eta}},
\end{eqnarray}
where the function $I_{0}(z)=\frac{1}{2\pi i}\oint e^{(z/2)(t+1/t)}t^{-1} dt$ is the modified Bessel function of the first kind.

Some simulations of Eq.~(\ref{key_rate}) assume that Alice and Bob use an infinite number of decoy intensity settings to estimate certain yields $p_{ZZ}(k_c,k_d|n_A,n_B)$ precisely. These quantities are given by
\begin{multline}\label{yieldsz}
p_{ZZ}(k_c,k_d|n_A,n_B)=\\
(1-p_d)\bigg[p_d(1-\sqrt{\eta})^{n_A+n_B}+q_{ZZ}(k_c,k_d|n_A,n_B)\bigg],
\end{multline}
where $q_{ZZ}(k_c,k_d|n_A,n_B)$ has the form
\begin{widetext}
\begin{multline}\label{yieldsz2}
q_{ZZ}(k_c,k_d|n_A,n_B)=\sum_{k=0}^{n_A}{n_A \choose k}\sum_{l=0}^{n_B}{n_B \choose l}\frac{\sqrt{\eta}^{k+l}(1-\sqrt{\eta})^{n_A+n_B-k-l}}{2^{k+l}k!l!}\sum_{m=0}^k{k \choose m}\sum_{p=0}^l{l \choose p}\sum_{q=\max{(0,m+p-l)}}^{\min{(k,m+p)}}{k \choose q}\\
{l \choose m+p-q}(m+p)!(k+l-m-p)!\cos^{m+q}{(\theta_{\rm A})}\cos^{m+2p-q}{(\theta_{\rm B})}\sin^{2k-m-q}{(\theta_{\rm A})}\sin^{2l-m-2p+q}{(\theta_{\rm B})}-(1-\sqrt{\eta})^{n_A+n_B}.
\end{multline}
\end{widetext}

Importantly, Eqs.~(\ref{yieldsz})-(\ref{yieldsz2}) show that the yields $p_{ZZ}(k_c,k_d|n_A,n_B)$ do not depend on the phase mismatch $\phi$, which results in Protocol~3 being quite robust against phase misalignment, as we discuss in the next section. 

\section*{Simulation results}

\begin{figure}[b]
  \scalebox{0.25}{\includegraphics{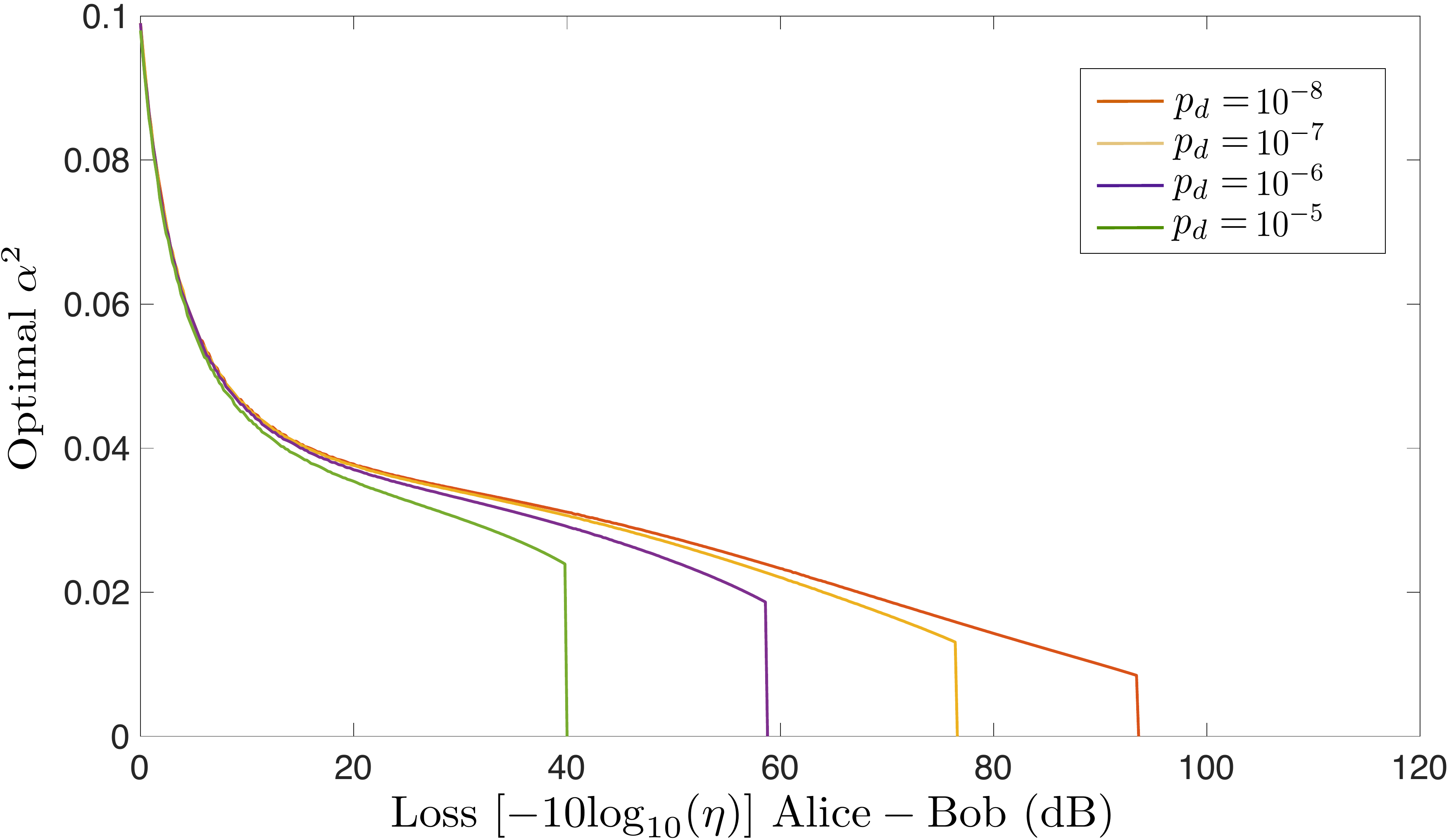}}
  \caption{Optimal values of the signal intensity $\alpha^2$ in Protocol~3 as a function of the overall loss between Alice and Bob. The different coloured lines identify different dark count rates, $p_d$, of the detectors in node $C$. These cases correspond to those shown in Fig.~1}\label{intensities}
\end{figure}
For completeness, in Fig.~\ref{intensities} in this Appendix we show the optimal values of the signal intensity, $\alpha^2$, as a function of the overall loss between Alice and Bob for the same cases considered in Fig.~1.

\begin{figure}[b]
  \scalebox{0.25}{\includegraphics{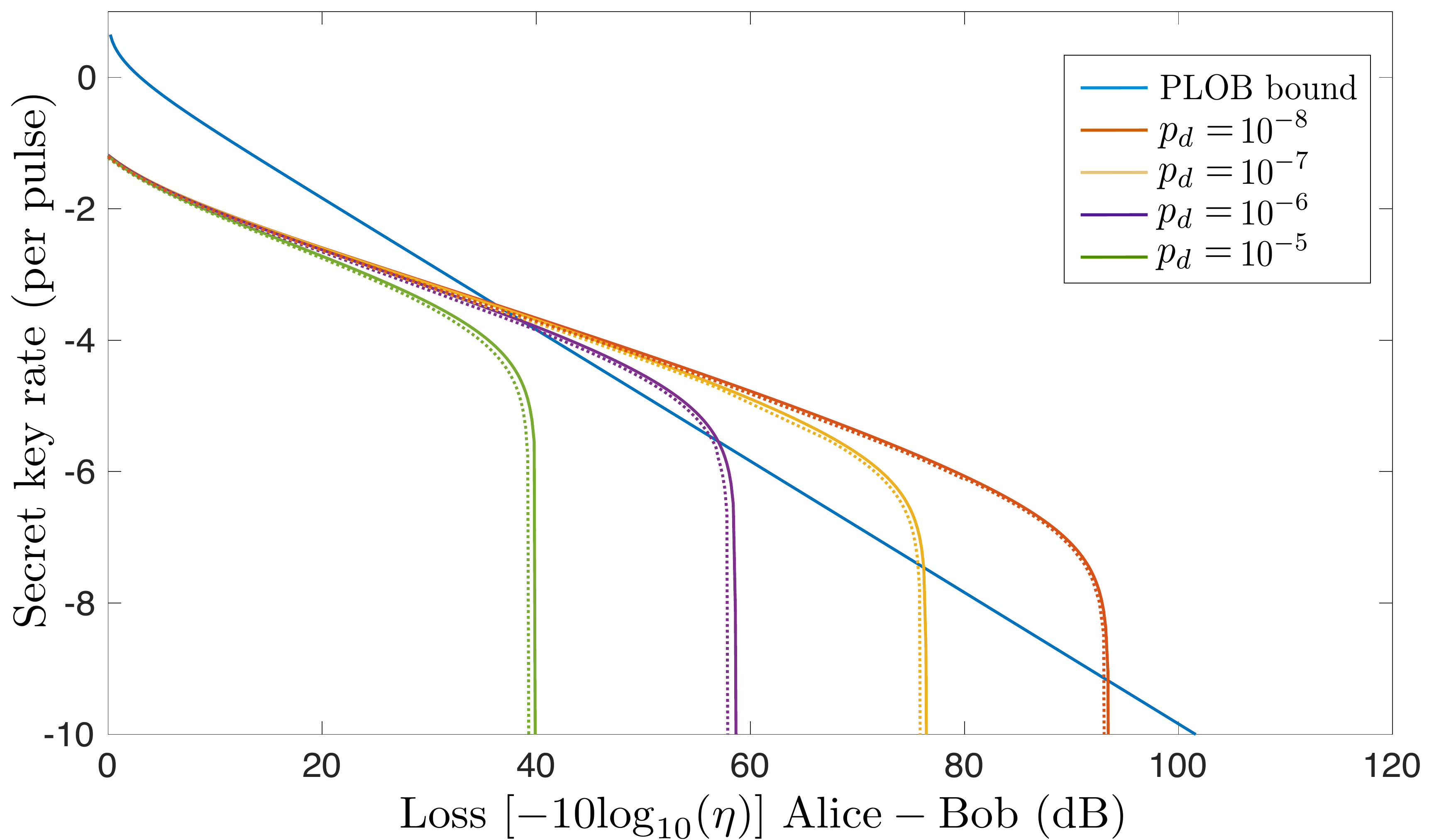}}
  \caption{Secret key rate (per pulse) in logarithmic scale for Protocol~3 as a function of the overall loss between Alice and Bob. The solid lines correspond to the cases illustrated in Fig.~1. They assume that Alice and Bob can estimate the yields $p_{ZZ}(k_c,k_d|2m_A+j, 2m_B+j)$ precisely for all $(m_A,m_B)\in{\cal S}_j$ for any $j\in \{0,1\}$. The dotted lines consider the practical scenario where Alice and Bob use only three decoy intensity settings each to estimate these quantities. Our simulation results demonstrate that three decoy intensity settings are enough to basically reproduce the performance of the asymptotic decoy state scenario studied in Fig.~1. 
    \label{fig_finite}}
\end{figure}

To plot the performance of Protocol~3 in Fig.~1, we consider, for simplicity, that the sets ${\cal S}_0$ and ${\cal S}_1$ are given by ${\cal S}_0=\{(0,0),(0,1),(1,0),(1,1)\}$ and ${\cal S}_1=\{ (0,0),(0,1),(1,0) \}$ respectively. Also, we assume that for all $(m_A,m_B)\in {\cal S}_j$, Alice and Bob can estimate the value of the yields $p_{ZZ}(k_c,k_d|2 m_A +j, 2m_B+j)$ precisely. That is, in this figure we suppose that the terms $p_{ZZ}^{\rm upp}(k_c,k_d|2 m_A +j, 2m_B+j)$ in Eq.~(\ref{phase_used}) satisfy $p_{ZZ}^{\rm upp}(k_c,k_d|2 m_A +j, 2m_B+j)=p_{ZZ}(k_c,k_d|2 m_A +j, 2m_B+j)$ with $p_{ZZ}(k_c,k_d|2 m_A +j, 2m_B+j)$ given by Eqs.~(\ref{yieldsz})-(\ref{yieldsz2}). This corresponds to the situation where Alice and Bob use infinite decoy intensity settings to estimate $p_{ZZ}^{\rm upp}(k_c,k_d|2 m_A +j, 2m_B+j)$ exactly. Moreover, we set the experimental parameters $\theta_{\rm A}=-\theta_{\rm B}=\arcsin{\sqrt{0.02}}$ which corresponds to a $2\%$ polarization misalignment, and we suppose that $\delta=0$. That is, for simplicity, in Fig.~1 we neglect the effect of the phase misalignment. This is motivated by the fact that, as we show below, Protocol~3 is actually quite robust to phase mismatch.

Fig.~\ref{fig_finite} considers the practical scenario where Alice and Bob use a finite number of decoy intensity settings to estimate $p_{ZZ}^{\rm upp}(k_c,k_d|2 m_A +j, 2m_B+j)$. Here, we suppose that each of them employs three decoy intensity settings, which, for simplicity, we set equal to $\beta_A^2, \beta_B^2\in\{0,0.001,0.1\}$. When the dark count rate, $p_d$, of the detectors at node $C$ is $p_d=10^{-6}$ ($p_d=10^{-8}$) we fine-tune a bit the decoy intensity choices, $\beta_A^2$ and $\beta_B^2$, around these intensity values. We do so in the loss region near the cut-off points where the key rates drop down to zero, {\it i.e.}, when the overall loss between Alice and Bob lies in the interval $[57{\rm dB},58 {\rm dB}]$ ($[80{\rm dB},93 {\rm dB}]$). The estimation of $p_{ZZ}^{\rm upp}(k_c,k_d|2 m_A +j, 2m_B+j)$ is done by solving the linear program given by Eq.~(\ref{lp}). For this, we use the sets ${\cal S}_0$ and ${\cal S}_1$ defined above, and we select ${\mathcal S}_{\rm cut}$ as ${\mathcal S}_{\rm cut}=\{(n_A,n_B) | n_A,n_B \in \mathbb{N}_0, n_A\le M_{\rm cut}, n_B \le M_{\rm cut} \}$ for a prefixed value $M_{\rm cut}$ that we choose equal to $10$. Also, we consider the same experimental parameters assumed in Fig.~1. We solve Eq.~(\ref{lp}) by using the linear programming solver MOSEK~\cite{mosek}, together with the parser YALMIP~\cite{yalmip}. 

Importantly, Fig.~\ref{fig_finite} demonstrates that only three decoy intensity settings are enough to basically reproduce the same performance of the asymptotic decoy state scenario illustrated in Fig.~1. This result is not surprising. Indeed, it can be shown that the most relevant terms $p_{ZZ}^{\rm upp}(k_c,k_d|2 m_A +j, 2m_B+j)$ in Eq.~(\ref{phase_used}) are those where $2 m_A +j+2m_B+j\leq{}2$. And these terms can be estimated very tightly already with three decoy intensities. Of course, the more decoy intensity settings Alice and Bob use, the more terms $p_{ZZ}(k_c,k_d|2 m_A +j, 2m_B+j)$ they can upper bound tightly, and thus the smaller is the estimated phase-error rate, and the higher is the resulting secret key rate. In addition, we remark that upper bounding many terms $p_{ZZ}(k_c,k_d|2 m_A +j, 2m_B+j)$ tightly also helps Alice and Bob to improve the secret key rate in a more subtle way. In particular, they can increase the amplitude $\alpha$ of the signal states, and thus increase the prefactor $p_{XX}(k_c,k_d)$ in Eq.~(\ref{qwe5}), without increasing much the residual parameters $\Delta_j$. However, as we will show below, these improvements are not too significant when compared to the case illustrated in Fig.~\ref{fig_finite}.   

\begin{figure}[b]
  \scalebox{0.25}{\includegraphics{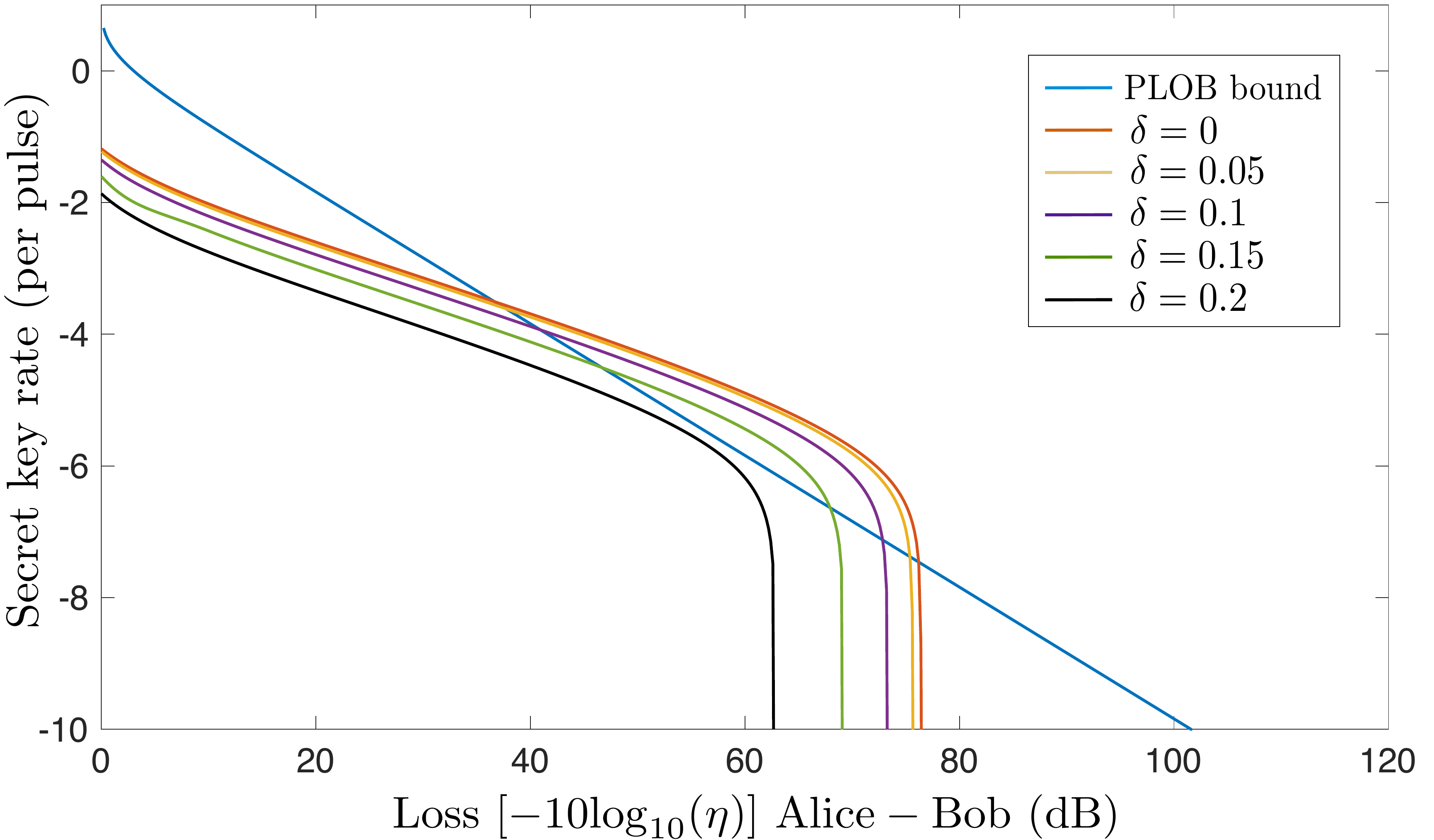}}
  \caption{Secret key rate (per pulse) in logarithmic scale for Protocol~3 as a function of the overall loss between Alice and Bob, for various values of the phase mismatch parameter $\delta\in\{0,0.05,0.1,0.15,0.2\}$. All other experimental parameters are equal to those considered in Fig.~1, and we set $p_d=10^{-7}$.  Our simulation results indicate that Protocol~3 is quite robust against phase mismatch. Indeed, when $\delta$ is about $5\%$ the results are almost indistinguishable from those of the ideal scenario with no phase mismatch. Also, even when $\delta$ is as high as $15\%$ one can still beat the PLOB bound. This is mainly because phase mismatch does not increase the phase-error rate but only affects the bit-error rate of the $X$-basis signals used for key generation.}
  \label{phase_mis}
\end{figure}

\begin{figure}[b]
  \scalebox{0.25}{\includegraphics{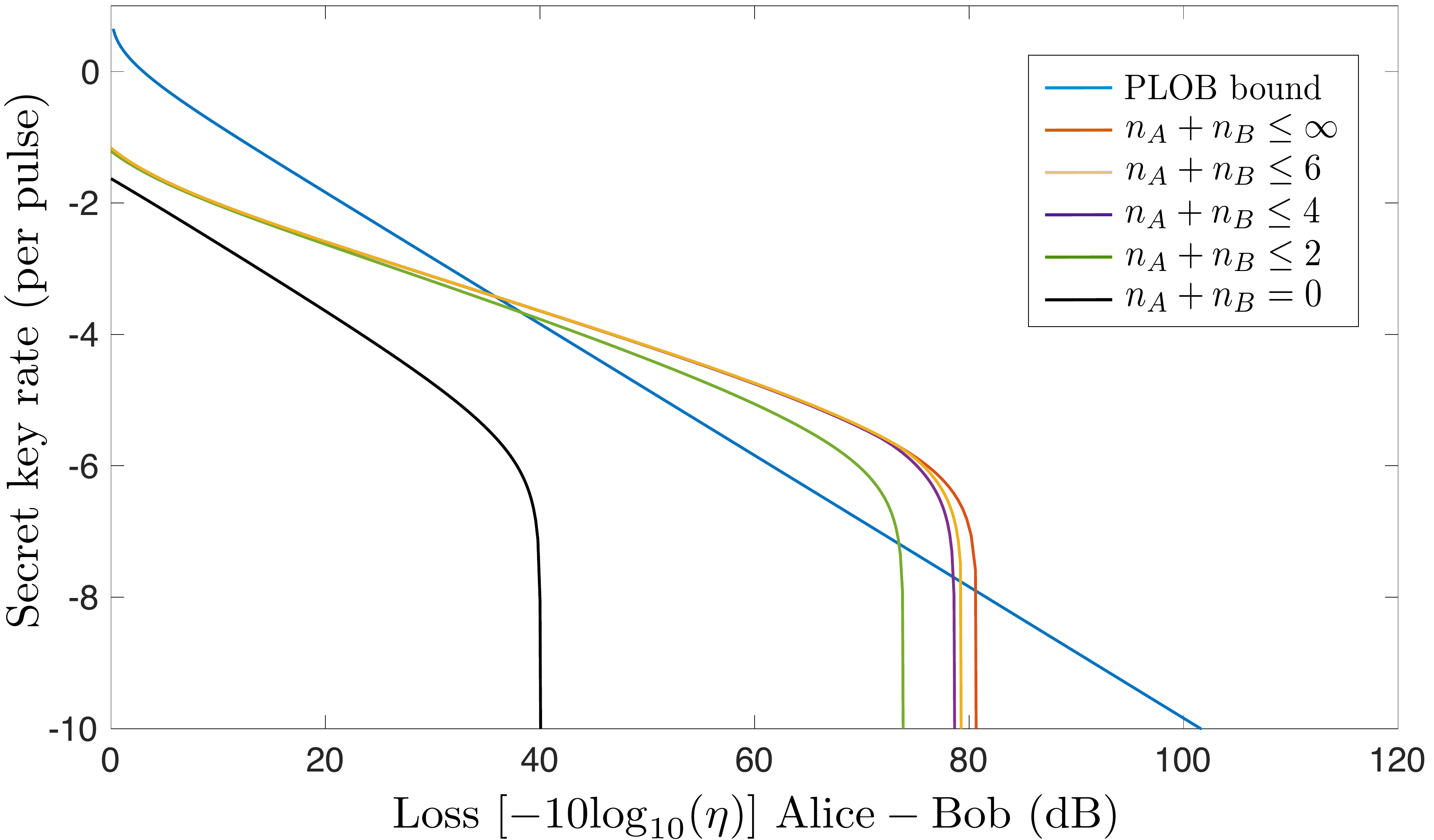}}
  \caption{Secret key rate (per pulse) in logarithmic scale for Protocol~3 as a function of the overall loss between Alice and Bob for different sets ${\cal S}_0$ and ${\cal S}_1$. More precisely, we consider that ${\cal S}_0=\{ (m_A,m_B) | 2 m_A+2 m_B \le N_{\rm max}  \}$ and ${\cal S}_1=\{ (m_A,m_B) | 2 m_A +1+2 m_B+1 \le N_{\rm max}  \}$ for given $N_{\rm max}$. Moreover, for simplicity, we assume that Alice and Bob can estimate the yields $p_{ZZ}(k_c,k_d|2 m_A +j, 2m_B+j)$, with $2 m_A +j+ 2m_B+j \le N_{\rm max}$, precisely. In this figure we use the same experimental parameters employed in Fig.~1, and we set $p_d=10^{-7}$. Our simulation results indicate that the case $N_{\rm max}=4$ already delivers a secret key rate very close to the ideal scenario where one evaluates Eq.~(\ref{phase}) exactly.}
  \label{last}
\end{figure}

So far, in our simulations we have neglected the effect of the phase mismatch between Alice and Bob's signals. This effect is investigated in Fig.~\ref{phase_mis}. For illustrative purposes, in this figure we consider that $p_d=10^{-7}$, and we evaluate the resulting secret key rate for various values of the parameter $\delta\in\{0,0.05,0.1,0.15,0.2\}$. The case $\delta=0$ corresponds to that shown in Fig.~1. The results illustrated in Fig.~\ref{phase_mis} confirm that Protocol~3 is actually quite robust against phase mismatch. Indeed, when the phase mismatch is about $5\%$ the secret key rate is almost indistinguishable from that of the ideal scenario with no phase mismatch. Even when the phase mismatch is as high as $15\%$ one can still beat the PLOB bound. As we have mentioned briefly earlier, this is mainly because phase mismatch does not affect the yields $p_{ZZ}(k_c,k_d|n_A,n_B)$ of the $Z$-basis signals, and, as a consequence of that, it does not increase the phase-error rate of the $X$-basis signals used for key generation. The phase mismatch only affects the bit-error rate of the $X$-basis signals. As a result, if the phase mismatch is not too large, it turns out that its effect on the secret key rate is relatively small, as shown in Fig.~\ref{phase_mis}.

To conclude this section, we investigate the effect that the sets ${\cal S}_0$ and ${\cal S}_1$ have on the secret key rate. For this, we now consider that these sets satisfy ${\cal S}_0=\{ (m_A,m_B) | 2 m_A+2 m_B \le N_{\rm max}  \}$ and ${\cal S}_1=\{ (m_A,m_B) | 2 m_A +1+2 m_B+1 \le N_{\rm max}  \}$. That is, we suppose that Alice and Bob nontrivially upper bound only those yields $p_{ZZ}(k_c,k_d|2 m_A +j, 2m_B+j)$, with $2 m_A +j+ 2m_B+j \le N_{\rm max}$, for certain number $N_{\rm max}$, which, without loss of generality, we shall consider is an even number because $2m_A+j+2m_B+j$ is also even. In addition, for simplicity, we shall consider that Alice and Bob can estimate the yields $p_{ZZ}(k_c,k_d|2 m_A +j, 2m_B+j)$, with $(m_A,m_B)\in {\cal S}_j$, precisely. We remark, however, that similar results can be obtained as well when they use a finite number of decoy intensity settings, as we have shown in Fig.~\ref{fig_finite}.

The simulation results are shown in Fig.~\ref{last}. In this figure we use the same experimental parameters employed in Fig.~1, and we set $p_d=10^{-7}$. The case $N_{\rm max}=0$ corresponds to the situation where Alice and Bob nontrivially upper bound only the yield $p_{ZZ}(k_c,k_d|0,0)$. At first sight, it might seem surprising that this is already enough to obtain a positive key rate over about 40 dB loss between  Alice and Bob, as shown in Fig.~\ref{last}. However, we remind the readers that although the main contribution to the secure key comes from the single-photon events, {\it i.e.}, when $(n_A,n_B)=(0,1)$ or $(1,0)$, such single-photon events do not contribute to the phase error rate. So, even though $N_{\rm max}=0$, the phase error can still be bounded. Fig.~\ref{last} also includes the extreme case where $N_{\rm max}=\infty$, which corresponds to the situation where Alice and Bob can evaluate Eq.~(\ref{phase}) precisely. For this, we first consider a lower bound on the secret key rate for a relatively high number $N_{\rm max}$, say $N_{\rm max}=12$. As mentioned above, here we suppose that Alice and Bob can estimate all the yields in ${\cal S}_0$ and ${\cal S}_1$ precisely. Then, we evaluate an upper bound on the secret key rate by setting again $N_{\rm max}=12$ but now assuming that the residual terms $\Delta_j$ in Eq.~(\ref{phase_used}) are equal to zero. In this scenario, since such residual terms are so extremely small, it turns out that with the resolution of Fig.~\ref{last} both the lower and upper bound on the secret key rate basically overlap each other. This means that they are also indistinguishable from the case $N_{\rm max}=\infty$. Importantly, Fig.~\ref{last} suggests that considering only small photon numbers, until to say $N_{\rm max}=4$, is enough to obtain a secret key rate close to the ideal scenario $N_{\rm max}=\infty$.

\end{document}